\documentclass{llncs}
\usepackage{amsmath, amssymb, color, times, graphics, graphicx, verbatim} 
\usepackage{geometry}

\newtheorem{prob}{Problem}
\newtheorem{fact}{Fact}
\newcommand{\dtime}{d_{\mathrm{time}}}
\renewcommand{\epsilon}{\varepsilon}
\newcommand{\bigh}{{\mathcal H}}
\newcommand{\bigp}{{\mathcal P}}
\newcommand{\bigb}{{\mathcal B}}

\title{Moving Walkways, Escalators, and Elevators\thanks{This work was partially supported by the {\em Actions de Recherche Concert\'ees (ARC)\,} fund of the {\em Communaut\'e fran\c{c}aise de Belgique}.}}
\date{}

\author{
Jean Cardinal \inst{1} 
\and S\'ebastien Collette \inst{1}\thanks{Aspirant du FNRS.}
\and Ferran Hurtado \inst{2}\thanks{Research of Ferran Hurtado partially supported by projects
MEC MTM2006-01267 and Gen. Cat. 2005SGR00692.}
\and \\ Stefan Langerman \inst{1}\thanks{Chercheur Qualifi\'e du FNRS.} 
\and Bel\'en Palop \inst{3}\thanks{Research of Bel\'en Palop partially supported by project VA031B06 and HP2005-0137.}
}

\institute{Computer Science Department\\
Universit\'e Libre de Bruxelles CP 212\\
B-1050 Brussels, Belgium\\
\email{\{jcardin,secollet,slanger\}@ulb.ac.be} \and
Departament de Matem\`atica Aplicada II\\
Universitat Polit\`ecnica de Catalunya
Jordi Girona, 1--3\\    
E-08034 Barcelona, Spain\\    
\email{Ferran.Hurtado@upc.edu} \and
Departamento de Inform\'atica\\
Universidad de Valladolid, Spain\\
\email{b.palop@infor.uva.es}}

\begin{document}
\maketitle
\sloppy

\begin{abstract}
We consider algorithms for finding the optimal location of a simple transportation device, that we call a {\em moving walkway}, consisting of a pair of points in the plane between which the travel speed is high. More specifically, one can travel from one endpoint of the walkway to the other at speed $v>1$, but can only travel at unit speed between any other pair of points. The travel time between any two points in the plane is the minimum between the actual geometric distance, and the time needed to go from one point to the other using the walkway. A location for a walkway is said to be optimal with respect to a given set of points if it minimizes the maximum travel time between any two points of the set. We give a simple linear-time algorithm for finding an optimal location in the case where the points are on a line. We also give an $\Omega (n\log n)$ lower bound for the problem of computing the travel time diameter of a set of points on a line with respect to a given walkway. Then we describe a $O(n\log n)$ algorithm for locating a walkway with the additional restriction that the walkway must be horizontal. This algorithm is based on a recent generic method for solving quasiconvex programs with implicitly defined constraints. It is used in a $(1+\varepsilon )$-approximation algorithm for optimal location of a walkway of arbitrary orientation. Finally, we consider several variants of the transportation model, including {\em escalators}, used for traveling between two floors of a building, and {\em elevators}, which are strictly vertical.
\end{abstract}

\section{Introduction}

Finding the optimal location of a facility modeled as a geometric object is an old, well-studied topic in computational geometry. One of the simplest instances of such a problem is finding the smallest disk covering a given set of points, which can be done in linear time, or its min-sum counterpart, the Fermat-Weber problem, which is difficult to solve exactly in usual models of computation. Surveys on both combinatorial and geometric facility location problems can be found in~\cite{Drez96,Drez01,RT90}. The field is also tightly related to the theoretical notions of center, median and depth of a set of points~\cite{Greg}.

The optimality of a location is often defined with respect to a distance measure.
Recently, an interest has emerged in the definition and analysis of time distances, measuring the travel time between points in the plane given a geometric transportation facility, such as a highway or a geometric network made of line segments~\cite{AHIK03,AHP04,AHP05,AAP04,GSW07}. In general, the transportation facility is a subset of the plane where the travel speed is high, and the travel time is naturally defined as the minimum time needed to reach a point. Primary concerns in previous works on these notions include in particular efficient constructions of the Voronoi diagrams for these distances.\\

In this work, we consider optimal location of geometric transportation facilities on a line and in the plane. The transportation facilities we define are probably the simplest one can think about, and consist of a pair of points between which one can travel faster than anywhere else in the plane. This simple definition can be used to model for instance moving walkways in airports, shuttle connections or tunnels in a city. For simplicity, we refer to these as {\em moving walkways}. Given a set of points and a moving walkway, we define the travel time diameter as the maximum travel time between two points in the set. A moving walkway is said to be optimal if it minimizes the travel time diameter.\\

Moving walkways on a line are defined and analyzed in section~\ref{sec:1D}. We first provide an $\Omega (n\log n)$ lower bound on the computation of the travel-time diameter for a given walkway in the algebraic computation tree model. Then we give a simple linear-time algorithm for finding the optimal moving walkway on a line.

Section~\ref{sec:roa} summarizes some previously known results on LP-type and quasiconvex programs. 

In section~\ref{sec:2D}, we define moving walkways in the plane, and provide an $O(n\log n)$ deterministic algorithm for the diameter. We show, using Chan's technique for implicit quasiconvex programs, that this implies an $O(n\log n)$ algorithm for finding an optimal horizontal walkway.
We then derive an algorithm for computing a $(1+\epsilon)$-approximation of an optimal moving walkway with arbitrary orientation and constant speed in time $O(\epsilon^{-1}n\log n)$.

Section~\ref{sec:EaE} contains a discussion on how the transportation model can be restricted to yield the {\em escalator} and the {\em elevator} models, and on the complexity of the corresponding location algorithms. We also give an algorithm for computing the travel time diameter of a set of red and blue points with respect to $k$ elevators.

\paragraph{Related works.} We are not aware of many previous contributions on the topic of optimal transportation facility location based on time distances.
 
In~\cite{minmaxmin06}, we presented a family of geometric optimization problems in which the objective function was in a min-max-min form. As an example,
we proposed the problem of finding the location of a moving walkway on a line such that the maximum travel time between $n$ given pairs of source-destination points is minimized. The current paper presents extensions and generalizations of this result. 

Recently, an effort has been made by several authors to solve the problem of locating a highway that minimizes the maximum travel time among a set of points~\cite{AAA07}. In this work, a highway is modeled as a line in the plane on which the travel speed is higher, and customers can enter and exit the highway at any point on the line. In particular, they give an algorithm for the case where the travel speed on the highway is infinite, and another algorithm for the case where the highway is restricted to be vertical. A $O(n^2\log n)$ algorithm for the general problem is also given. Various other models are studied by Korman and Tokuyama~\cite{KT07}. The goal of this work is similar, but the transportation model that we study is different.

\section{Moving walkways on a line}
\label{sec:1D}

\begin{figure}
\begin{center}
\includegraphics[scale=.8]{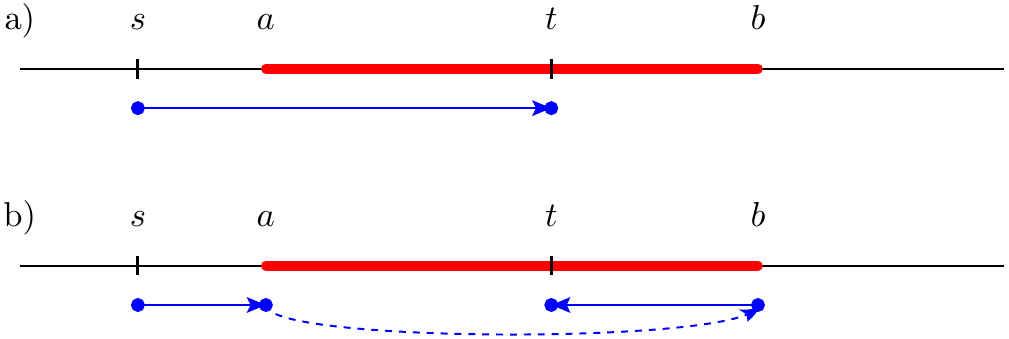}
\end{center}
\caption{\label{1Dww}Moving walkways on a line. The travel time between two points is the minimum between a) the direct distance,
b) the travel time using the walkway.}
\end{figure}

Suppose a new moving walkway is to be installed in a long corridor, for instance in the concourse of an airport. The moving walkway is used by customers to go from source to destination points in the corridor. We model the corridor as the real line, source and destination points as real numbers, and the moving walkway by an interval $[a,b]$ on the line. We denote by $v>1$ the speed on the moving walkway, and assume that the speed outside the moving walkway is 1. Customers can only enter the walkway at point $a$, step down of it at point $b$, and always follow the shortest path. This is illustrated in figure~\ref{1Dww}.

The travel time between a point $s$ of the line and any other point $t>s$, given a walkway $[a,b]$, is a time distance~\cite{AHP04,AAP04}, defined as 
follows:
$$
\dtime (s, t, a, b) = \min\{ t-s,\vert s - a\vert + \vert t -b\vert + \frac{1}{v}(b-a)\} .
$$
Note that by assuming $t>s$ we implicitly make the moving walkway bidirectional: for going from $t$ to $s$, we enter the walkway at point $b$ and exit at point $a$, which makes the travel time a symmetric distance function.

We consider the computation of the travel-time diameter and the optimal location of such a moving walkway.

\subsection{Travel time diameter}

A first problem is, given a set of points and a walkway $[a,b]$, to compute the travel time diameter, defined as 
the maximum travel time between two points of the set.

\begin{prob}
Given a set $P$ of $n$ real numbers, a real number $v>1$, and a pair $(a, b)$ of real numbers, find 
\begin{equation*}
\max_{s, t\in P : s\leq t} \dtime (s, t, a, b).
\end{equation*}
\end{prob}

\begin{theorem}
The travel time diameter for a moving walkway on a line can be computed in $O(n\log n)$ time.
\end{theorem}
\begin{proof}
Without loss of generality, we assume that $\max \{ x\in P \}  = 1$ and $\min \{ x\in P\} =0$.
We distinguish the four intervals $[0,a]$, $(a,(a+b)/2]$, $((a+b)/2,b]$ and $(b,1]$, and call these I, II, III and IV, respectively. 

We remark that if a pair $s, t$ is not separated by the point $(a+b)/2$, then the travel time between
$s$ and $t$ is $| t - s |$, i.e. we never use the walkway.
From this, the pairs of points between which the moving walkway is used for traveling can only be of four types:
II-III, I-III, II-IV and I-IV. For all the other cases, the travel time is equal to the maximum absolute difference within the 
corresponding sets of points.\\

\begin{figure}
\begin{center}
\includegraphics[scale=.6]{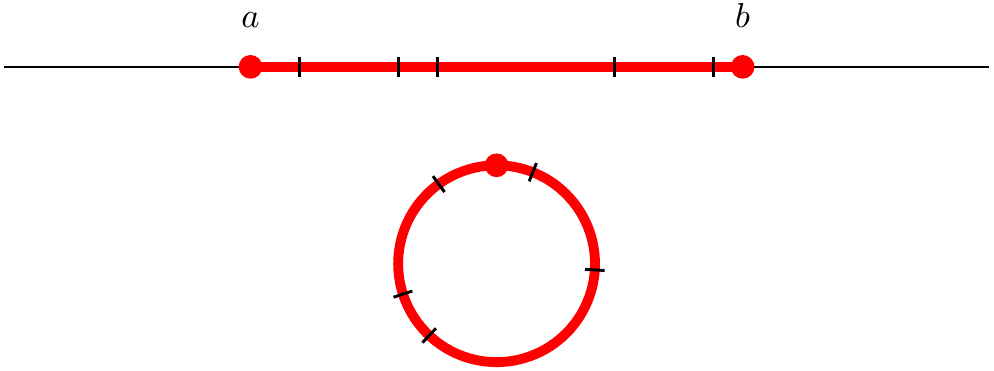}
\end{center}
\caption{\label{illdiam}Computing the travel time diameter for points in the interval $[a,b]$ amounts to computing the angular diameter of points on
a circle (here $v=+\infty$).}
\end{figure}

For the case II-III the computation of the diameter is essentially equivalent to computing the maximum angle between two 
points on a circle (see figure~\ref{illdiam}). This can be achieved in linear time once the input is sorted using
rotating calipers~\cite{rc}.

The case I-III is solved by computing the farthest point from $a$ on the same circle, which can also be achieved by scanning the 
sorted array. The case II-IV is symmetric to the case I-III. Finally we know that all pairs of type I-IV use the walkway, so the largest
travel time is $\dtime (0,1,a,b)$, which can be found in constant time.
\qed\end{proof}

\begin{theorem}
\label{diamlb}
There is an $\Omega(n\log n)$ lower bound on the computation of the travel time diameter for a moving walkway on a line in
the algebraic computation tree model, even when $v=+\infty$. 
\end{theorem}
\begin{proof}
The proof of the lower bound is by reduction from the problem of set disjointness, consisting of checking whether $A\cap B=\emptyset$ for
two sets of numbers $A$ and $B$. The proof is similar to the proof of the same lower bound for the problem of
computing the Euclidean diameter of a set of points. 

We let $v=+\infty$, $a=0$, $b=2$ and scale the elements of $A$ and $B$ so that they are all strictly smaller than 1. We then let 
$$
P= A \cup \{ x+1:x\in B\}.
$$ 
Then the travel diameter of the points in $P$ with respect to the moving walkway $a,b$ is equal to 1 iff $A\cap B\not= \emptyset$.
Note that since 1 is an upper bound on the travel time diameter, the lower bound also holds for the decision version of the problem.
\qed\end{proof}

\subsection{Minimum travel time diameter location}

We now consider the following moving walkway location problem.

\begin{prob}
\label{apmww}
Given a set $P$ of $n$ real numbers, a real number $v>1$, find a moving walkway $[a,b]$ of speed $v$ that 
minimizes the travel time diameter of $P$:
\begin{equation*}
\min_{a,b} \max_{s, t\in P : s\leq t} \dtime (s, t, a, b).
\end{equation*}
\end{prob}

There is a very simple deterministic algorithm solving this problem in $O(n)$ time. It is a surprising observation that we are able to find a moving walkway that minimizes the diameter faster than we are able to compute the diameter for a given walkway. This observation has already been made for the construction of optimal vertical highways~\cite{AAA07}. We assume w.l.o.g. that $\min\{ x\in P\} =0$ and $\max\{ x\in P\} =1$, and denote by $\text{OPT}$ the optimum value $\min_{a,b} \max_{s,t\in P :s\leq t} \dtime (s, t, a,b)$.

\begin{lemma}\label{upperbound}
$$
OPT \leq \frac{v}{2v-1}.
$$
\end{lemma}
\begin{proof}
We define a moving walkway by $a=(1 - {v\over 2v-1})/2 = {v-1\over 4v-2}$ and $b=({v\over 2v-1} + 1)/2 = {3v-1\over 4v-2}$.
Let $c_1 = 1-{v\over 2v-1}$ be the symmetric of point $0$ with respect to $a$. Similarly let $c_2={v\over 2v-1}$ be the symmetric of $1$ with respect to $b$. Then going from any point in $[0,c_1]$ to a point in $[c_2,1]$ using the walkway cannot take more time than going from $0$ to $1$. 

To go from $0$ to $1$, the time needed is $$a+(1-b)+{1\over v}(b-a)={v\over 2v-1}$$ For every pair in $[0,c_2]$, the time distance is at most ${v\over 2v-1}$ (the length of the interval). The same observation holds for every pair in $[c_1,1]$.
\qed\end{proof}

In what follows, we assume we are given a set of points $P$ and we try to find properties of its optimal moving walkway $[a,b]$. Let $s$ be the smallest point in $P$ such that, to go from $0$ to $s$, it is better to use the moving walkway. We know that $s \le b$, as to go from 0 to any point larger than or equal to $b$ it is always better to use the moving walkway. Symmetrically, let $r$ be the largest point in $P$ such that to go from $1$ to $r$, it is better to use the moving walkway; we know that $r \ge a$.

The following facts are easy to verify:

\begin{fact}\label{myfact1}
For every pair $(p,q)$ enclosing the walkway, i.e., $p\le a \le b \le q$, the walkway is used.
\end{fact}
\begin{fact}\label{myfact2}
If we have $p\leq p'\leq a$, and the walkway is used for the pair $(p',q)$, it is also used for the pair $(p,q)$. 
\end{fact}
\begin{fact}\label{myfact3}
If the walkway is used for the pair $(p,q)$ (with $p\leq q$), and $p'$ is between $p$ and $a$, and $q'$ is between $q$ and $b$, then the walkway is used
for the pair $(p',q')$.
\end{fact}

\begin{lemma}\label{whenneeded}
Let $P$ be a set of points and $[a,b]$ an optimal moving walkway. The moving walkway is used only for pairs $(p,q)$ where $p \in [0,r]$ and $q \in [s,1]$. For every other pair, the facility is not needed.
\end{lemma}
\begin{proof}
We first show that for a pair $(p,q)$ in the interval $[0,s)$, the facility is never used. We distinguish three cases.
\begin{enumerate}
\item if $(p,q)$ belongs to $[0,a]$, the walkway is trivially not used.

\item if $p\le a$ and $q$ belongs to $[a,s)$, then using the walkway for going from $p$ to $q$ implies that we should use it for going from $0$ to $q$ (see Fact~\ref{myfact2}). This contradicts the definition of $s$.

\item if both $p$ and $q$ belong to $[a,s)$, then using the walkway for going from $p$ to $q$ implies that we should use it also for going from $p'$ to $q$, where $p'$ is the symmetric of $p$ with respect to $a$. We are then back to case 2, leading to a contradiction. 
\end{enumerate}
In conclusion, the walkway is never used for pairs in the interval $[0,s)$. The same reasoning holds for pairs in $(r,1]$.
\qed\end{proof}

\begin{lemma}~\label{middle}
The optimal moving walkway is such that $a={r \over 2}$ and $b={s+1\over 2}$.
\end{lemma}
\begin{proof}
We have not proved that \emph{all} pairs $(p,q)$ with $p \in [0,r]$ and $q \in [s,1]$ require the use of the moving walkway. What we know however is that the moving walkway will be used for the pairs $(0,1)$, $(0,s)$ and $(r,1)$. What we aim to show is that finding the optimal moving walkway is equivalent to finding the optimal moving walkway for these pairs. 

First note that the optimal walkway for $(0,1)$, $(0,s)$ and $(r,1)$ is such that $a={r \over 2}$ and $b={s+1\over 2}$. Moreover, it implies that $\dtime (0,1,a,b)=\dtime (0,s,a,b)=\dtime (r,1,a,b)$. We also have $\dtime (0,1,a,b) \geq \dtime (r,s,a,b)$, meaning that we also optimize the distance for $(r,s)$.

For every pair with $p \in [0,r]$ and $q \in [s,1]$, we have one of the following:
\begin{itemize}
\item if $p\le a$ and $q\ge b$, the walkway is always used (Fact~\ref{myfact1}),
\item if $p\le a$ and $q\le b$, the walkway is always used since it is used for $(0,s)$ (Fact~\ref{myfact3}),
\item symmetrically, if $p\ge a$ and $q\ge b$, the walkway is used since it is used for $(r,1)$,
\item if $p\ge a$ and $q\le b$ and the walkway is not used, then the time distance for $(p,q)$ is no greater than the shortest distance from $p$ to $q$ \emph{using} the walkway, which is no greater than the time distance for $(r,s)$. If it is used, it is no greater than the time distance for $(r,s)$.
\end{itemize}
So, in every case, the time distance is upper bounded by that of $(0,1)$, $(0,s)$, $(r,1)$ or $(r,s)$, the maximum of which is minimized with the solution described above. 
\qed\end{proof}

\noindent This lemma showed that the optimal moving walkway is uniquely determined by the points $r$ and $s$. 

\begin{lemma}\label{r_and_s}
Let 
\begin{itemize}
\item $r_{1}$ be the largest point of $P$ whose value is less than or equal to $1 - {v \over 2v-1}$
\item $s_{1}$ be the smallest point of $P$ whose value is more than or equal to $r_{1}(v-1)+v+1\over 3v-1$
\item $r_{2}$ be the largest point of $P$ whose value is less than or equal to $(1-v)(s_{2}+1)\over 1-3v$
\item $s_{2}$ be the smallest point of $P$ whose value is more than or equal to ${v\over 2v-1}$.
\end{itemize}
\noindent Then, either $r=r_{1}$ and $s=s_{1}$ or $r=r_{2}$ and $s=s_{2}$.
\end{lemma}

\begin{proof}
By Lemma~\ref{upperbound} and Lemma~\ref{middle} we know that either $r\le 1 - {v \over 2v-1}$ or $s\ge {v\over 2v-1}$, for otherwise the walkway $(a=r/2,b={s+1\over 2})$ leads to a diameter greater than $\frac{v}{2v-1}$. We treat the first case, the other one being symmetric.

If $r\le 1 - {v \over 2v-1}$, we prove that $r=r_{1}$ and $s=s_{1}$.
We proceed in two steps: first we will show that $r=r_{1}$ is optimal, and then prove that $s=s_{1}$ independently of the choice of $r$.

By contradiction, imagine that the optimal $r$ is not $r_{1}$. Because we assumed that $r\le 1 - {v \over 2v-1}$, this means that there exists a point $q$ between $r$ and $1 - {v \over 2v-1}$, as $r$ is not the largest. Since $q>r$, and $r$ is, by definition, the largest point for which we should use the walkway to go to $1$, we know that the walkway is \emph{not} used from $q$ to $1$, leading to a diameter greater than $v\over 2v-1$, and contradicting Lemma~\ref{upperbound}. To conclude, we know that in the case where $r\le 1 - {v \over 2v-1}$, $r$ is the largest point of $P$ satisfying that constraint. 

Now we prove that $s=s_{1}$. By definition of $s$, it is the first point in $P$ such that, to go from $0$ to $s$, it is better to use the walkway. Thus we obtain the following equation: 

$$s\ge a+{b-a\over v} + b-s={r\over 2} + \left({1+s\over 2}-{r\over 2}\right){1\over v} + {{1-s\over 2}}$$
$$s\ge{r(v-1)+v+1\over 3v-1}$$

By definition, $s$ is the \emph{smallest} point in $P$ satisfying that constraint, and thus, as claimed in the statement of the Lemma, $s=s_{1}$.
\qed\end{proof}







\begin{theorem}
Problem~\ref{apmww} can be solved in $O(n)$ deterministic time.
\end{theorem}
\begin{proof}
By Lemma~\ref{r_and_s}, we know that there are two possibilities. So first, we determine the values of $r_{1}$, $r_{2}$, $s_{1}$ and $s_{2}$ in linear time (we scan the set twice). By Lemma~\ref{middle}, the optimal moving walkway is given by 
$a={r \over 2}$ and $b={s+1\over 2}$, so we have two candidates.

By Lemma~\ref{whenneeded} and the arguments developed in the proof of Lemma~\ref{middle}, we know that the largest time distance for pairs $(p,q)$ with $p \in [0,r]$ and $q \in [s,1]$ is achieved by one of the pairs $(0,1)$, $(0,s)$, $(r,1)$ or $(r,s)$. The largest time distance for pairs where both points lie in $[r,s]$ only depends on the smallest and largest point in that ranges and can also be found by scanning the set. For the pairs in $[0,r]$ or in $[s,1]$, the worst distance achieved is $\max{(r,1-s)}$. 

In other words, we compute the worst time distance for both candidates (for pairs using and not using the walkway), and keep the best. 
\qed\end{proof}

\begin{figure}
\begin{center}
\includegraphics[scale=.8]{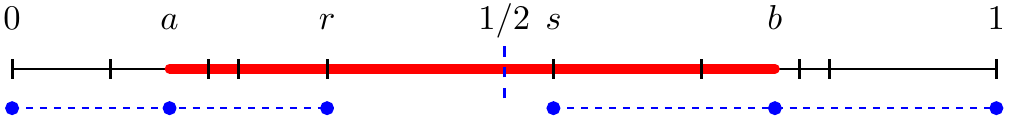}
\end{center}
\caption{\label{belen}Illustration of the moving walkway location algorithm on a line.}
\end{figure}

\section{Randomized optimization algorithms}
\label{sec:roa}

In this section, we briefly recall useful results on generic optimization problems. These results will be used in the next section on moving walkway
location in the plane.\\

The problems we consider are constrained minimization problem of the form 
\begin{eqnarray*}
& \min_{x,y} &  y \\
& \text{subject\ to} & y\geq f_i(x)\ \forall i=1,2,\ldots ,n,
\end{eqnarray*}
where $x\in{\mathbb R}^{d-1}$, $y\in{\mathbb R}$ and $\{f_i\}$ is a collection of continuous functions of the form $f_i:{\mathbb R}^{d-1}\mapsto {\mathbb R}$. In order to have a unique minimum, we can define the $\min$ operator as a lexicographic minimum on the vector made of the value $y$ and the coordinates of $x$. Geometrically, the problem can be interpreted as seeking a lowest point on the upper envelope of a set of $(d-1)$-dimensional surfaces $f_i(x)$. If these surfaces are hyperplanes, then we have a linear programming problem. If they are defined by convex functions, then we have a convex programming problem. In both cases, there exist efficient randomized algorithms running in $O(n)$ expected time.\\

In fact these algorithms do not apply solely to linear and convex programming problems, but to a well-defined family
of so-called LP-type problems~\cite{LPtype}. 
An abstract $d$-dimensional LP-type problem is a collection of constraints $\bigh$ and a function $w:2^{\bigh}\mapsto W$ associating a value in a totally ordered set $W$ to each subset of $\bigh$. The goal is to compute $w(\bigh )$. Moreover, $w$ has to satisfy the following, for any $H\subseteq\bigh$ and $h\in\bigh$:
\begin{enumerate}
\item $w(H) = w(B)$ for some $B\subseteq H$ of size at most $d$; $B$ is called a {\em basis} for $H$,
\item $w(H\cup\{h\})\geq w(H)$,
\item If $w(H)=w(B)$ and $B\subseteq H$, then $w(H\cup\{h\})=w(H)\Leftrightarrow w(B\cup\{h\})=w(B)$.
\end{enumerate}
Provided that $d$ is a constant and the computation model allows the execution of some elementary operations in constant time, then an abstract LP-type problem can be solved in $O(n)$ randomized expected time.
In the above problem, the set $W$ is the set $\mathbb R$, a constraint $h$ is a surface $f_i(x)$, and the value $w(H)$ of a set $H$ of constraint is the height of the lowest point above all the surfaces in $H$. If the functions $f_i$ are convex, the dimension of the LP-type problem is $d$, since an optimum $w(H)$ is defined by $d$ surfaces in $H$.\\

A function $f$ is called {\em quasiconvex} if all its (lower) levels are convex, i.e. if the set $\{ x : f(x)\leq t \}$ is convex for all value of $t$. Alternatively, a quasiconvex function is such that $f((x+x')/2)\leq\max \{f(x),f(x')\}$ for any pair $x,x'$. Quasiconvex programming is essentially the problem defined above in the case where all the functions $f_i$ are quasiconvex~\cite{qconvex,qcloc}. Amenta, Bern and Eppstein showed the following:
\begin{theorem}[\cite{abe99}]\label{qc99}
Any quasiconvex program of dimension $d$ forms an LP-type problem of dimension at most 2d+1.
\end{theorem}
As a direct consequence, any quasiconvex program with $n$ constraints can be solved in $O(n)$ randomized expected time.\\

Recently, Chan introduced a generic technique for solving some LP-type problems with a large number of constraints~\cite{Chan2}. He motivated the introduction of this method by the problem of computing the Tukey depth of a set of points. It has since been used in a number of other contributions, including a new algorithm from Morin for computing the Zonoid depth of a set of points~\cite{zonoid}, and a facility location algorithm from Eppstein and Wortman~\cite{stars}. We give the main result applied to quasiconvex programming, using notations from~\cite{qconvex}.

\begin{theorem}[\cite{Chan2}]\label{chan2005}
Let $\bigp$ be a set of input values and $\bigh$ a set of quasiconvex functions. If there exists a function $g: 2^{\bigp}\mapsto 2^{\bigh}$ such that:
\begin{enumerate}

\item the quasiconvex program defined by the constraints $g(B)$ with $B\subseteq\bigp$ can be solved in $O(1)$ time whenever $|B|=O(1)$,

\item \label{chandec} there exists an algorithm running in $O(D(|P|))$ time for deciding whether a given pair $(x,y)$ is a feasible solution for the program
defined by $P\subseteq\bigp$, i.e. whether $y\geq f(x)\ \forall f\in g(P)$, and such that $D(n)/n^{\epsilon}$ is monotone increasing for a constant 
$\epsilon >0$,

\item \label{chansplit} there exist two constants $\alpha<1$ and $r\in\mathbb N$ such that we can, in time $O(D(|P|))$, decompose the problem $P\subset\bigp$ into $r$ subproblems $P_1,P_2,\ldots ,P_r$ of size at most $\alpha |P|$ and such that $g(P)=\bigcup_{i=1}^r g(P_i)$ 

\end{enumerate}
then for any $P\in\bigp$, we can solve the quasiconvex program defined by $g(P)$ in randomized expected time $O(D(|P|))$.
\end{theorem}
So under some simple conditions, we are able to solve a quasiconvex program in essentially the same time as needed to decide whether a given solution
$(x,y)$ is feasible. This is achieved by a generalization of the standard randomized incremental algorithm in which the bases are evaluated recursively, and the satisfaction-violation test is handled by the algorithm provided by requirement~\ref{chandec}.

This is interesting in situations where the input is not an explicit list of constraints, but rather some data $P$ that can be mapped to a set of constraints $g(P)$ that is much larger. This yields what are called {\em implicit quasiconvex programs}. This is particularly useful for our minimum-diameter problems, since the diameter is defined by a pair of points in a $n$-point set, yielding $n\choose 2$ constraints. But if the above conditions are met and we can test the diameter for a particular solution in time $D(n)$, then we can find the minimum-diameter solution in time $O(D(n))$ as well.

\section{Moving walkways in the plane}
\label{sec:2D}

In this section, we generalize the previous moving walkway model to the plane. A moving walkway is defined by two points $a$ and $b$,
and customers can go from a point $s$ to a point $t$ either by going from $s$ to $t$ directly, or by entering the walkway at one of its endpoint and
exiting at the other. The walkway is parameterized by a real number $v$ so that the time to go from $a$ to $b$ is ${1\over v}d(a,b)$.
The distance measure $d(.,.)$ used in the definition of the travel time must be convex and satisfy the triangular inequality. For simplicity we
assume it is the Euclidean distance, keeping in mind that all the following can be generalized to other distance measures.
Note that the moving walkways are bidirectional, hence the travel time is a symmetric distance function:
$$
\dtime (s, t, a, b)=\min \{ d(s, t), d(s, a) + {1\over v}d(a,b) + d(b,t), d(s, b) + {1\over v}d(a,b) + d(t,a) \}.
$$
We first solve a location problem with $n$ source-destination pairs, under the restriction that the moving walkway must be horizontal.
Then we give an algorithm for computing the travel time diameter for a given moving walkway (not necessarily horizontal), and from this derive an algorithm for finding optimal horizontal walkways. We show that this algorithm can be used for obtaining an approximation algorithm for locating moving walkways with arbitrary orientation. 

\subsection{Horizontal moving walkway location for $n$ source-destination pairs}

As a preliminary, we consider the problem of minimizing the maximum travel time between a number of designated source-destination pairs, with the restriction that the moving walkway must be horizontal. We let $(p_x, p_y)$ denote the $x$ and $y$-coordinates of a point $p$. 

\begin{prob}
\label{2Dhmww}
Given $n$ pairs $\{(s_i, t_i)\}_{i=1}^n$ of points, and a real number $v>1$, 
find a horizontal moving walkway $[a^*,b^*]$ of speed $v$ minimizing the maximum travel time between a pair:
\begin{equation*}
\min_{a, b: a_y=b_y} \max_i \dtime (s_i, t_i, a, b),
\end{equation*}
\end{prob}

The following lemma states that if one travels from left to right, then the moving walkway, if it is used, will be used from left to right.
\begin{lemma}
Given a pair of points $a,b$ such that $a_y=b_y$ and $a_x<b_x$, and a pair of points $s,t$ such that $s_x<t_x$, we have:
$$
\dtime (s, t, a, b) = \min \{ d(s, t), d(s, a) + {1\over v}d(a,b) + d(b,t)\}.
$$
\end{lemma}
\begin{proof}
As it is the case on the line, pairs of points $s,t$ for which the walkway is useful are separated by the bisector of the segment $[a,b]$. Then
we can see that if we use the walkway from $b$ to $a$, using it from $a$ to $b$ can only reduce the travel time.
\qed\end{proof}

\begin{figure}
\begin{center}
\includegraphics[scale=1]{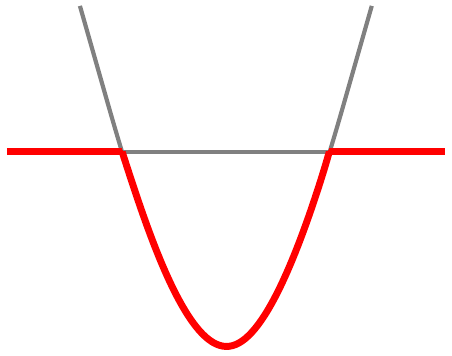}
\end{center}
\caption{\label{qconv}Functions defined as the minimum between a convex function and a constant are quasiconvex.}
\end{figure}

\begin{lemma}
Problem~\ref{2Dhmww} is LP-type.
\end{lemma}
\begin{proof}
We prove that the functions 
$$
f_i (a, b) = \dtime (s_i, t_i, a, b)
$$
are quasiconvex. From Theorem~\ref{qc99}, this implies that the problem is LP-type. 
The domain of these functions is $\mathbb R^4$, however we only consider their restriction to
the subspace of points $a,b$ such that $a_x\leq b_x$ and $a_y=b_y$, for which quasiconvexity 
is preserved. We also assume, without loss of generality, that ${s_i}_x \leq {t_i}_x$.

We define the following function $g_i$, defined on the same domain as $f_i$:
$$
g_i (a, b) = d(s_i, a) + {1\over v}d(a, b) + d(b, t_i).
$$
This function is convex, since it is a positively weighted sum of convex functions.

Now we have $f_i (a, b) = \min \{ d(s_i, t_i), g_i (a,b)\}$. The value $d(s_i, t_i)$ does not depend on $a$ and $b$, and
can therefore be considered as a constant. So $f_i$ is defined as the minimum between a constant and a convex function. All the levels
of such a function are necessarily convex (see figure~\ref{qconv}).
\qed\end{proof}

LP-type problems are known to be solvable in expected time linear in the number of constraints~\cite{LPtype}, which implies that Problem~\ref{2Dhmww}
is solvable in $O(n)$ expected time.

\begin{remark}
The restriction on the direction of the moving walkway is useful in making the functions $f_i$ quasiconvex. If this assumption
is lifted, then it is easy to show that the levels of the functions are not even always connected. There are examples such that when interpolating
between a pair of moving walkways $a,b$ and $a',b'$, the travel time starts to increase, then decrease afterwards because the customer
uses the walkway in the other direction. 
\end{remark}

\begin{remark}
In the above proof, we implicitly use the assumption on the computation model that constant-size subproblems can be solved in constant time. 
This is a standard assumption, but it can make the given algorithms difficult to implement in practice. 
\end{remark}

\subsection{Travel time diameter}

As a next step toward an algorithm for minimum travel time diameter location, we give an  $O(n\log n)$ time decision procedure for the travel time 
diameter given a moving walkway (not necessarily horizontal). Note that the travel time diameter for a moving walkway in the plane generalizes the travel time diameter for a moving walkway on a line, so the $O(n\log n)$ lower bound given in theorem~\ref{diamlb} still holds and our algorithm is optimal.

\begin{prob}
\label{2Ddiam}
Given a set $P$ of $n$ points, a real number $v>1$, and a pair of points $a, b$, find
\begin{equation*}
\max_{s, t\in P} \dtime (s, t, a, b).
\end{equation*}
\end{prob}

\begin{lemma}
The decision version of Problem~\ref{2Ddiam} can be solved in optimal $O(n\log n)$ deterministic time.
\end{lemma}
\begin{proof}

The decision problem is to check, for a given walkway $a,b$, whether the largest travel time $\max_{s, t\in P} \dtime (s, t, a, b)$ between two points 
of $P$ is not greater than a given value $y$.

We first observe that the pairs of points $s,t\in P$ such that $ \dtime (s,t,a,b) < d(s,t)$ are all separated by the bisector of the line segment $[ab]$.
If both points are on the same side of the bisector, then it is always more advantageous to go directly from $s$ to $t$ than to use the walkway. So as a first step, we can partition the point set in two subsets $R=\{ s\in P : d(s, a)\leq d(s,b) \}$ and $B=\{ t\in P: d(t, a)>d(t,b) \}$, say of red and blue points, according to which side of the bisector the points lie. Then we have :
$$
\max_{s,t\in P} \dtime (s,t,a,b) = \max\{ \max_{s, t\in R} d(s, t), \max_{s, t\in B} d(s, t), \max_{s\in R, t\in B} \dtime (s, t, a, b)\} .
$$

So by first computing, in time $O(n\log n)$, the Euclidean diameter of the red and blue sets, the travel time diameter problem boils down to computing the red-blue travel time diameter.
 
\begin{figure}
\begin{center}
\includegraphics[scale=.6]{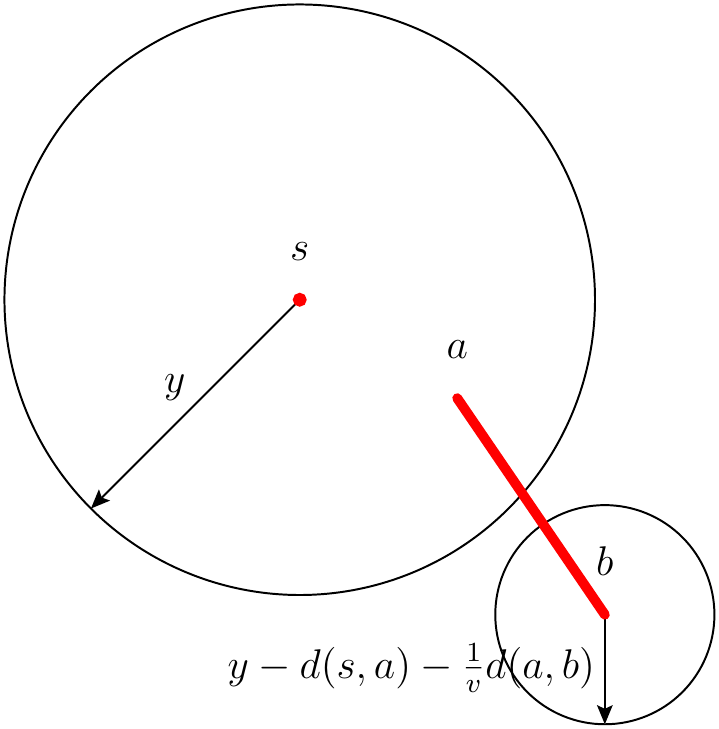}
\end{center}
\caption{\label{unit}The travel time disk $\bigb (s,y)$.}
\end{figure}

The red-blue travel time diameter is not greater than $y$ if and only if each point $t\in B$ is contained in the intersection of the {\em travel time disks} 
of the points in $R$, where the travel time disk $\bigb (s,y)$ of $s\in R$ is the set of points reachable in time $y$ from $s$. 
We observe that for a red-blue pair $s\in R, t\in B$, we have $\dtime (s, t, a, b) = \min\{ d(s,t), d(s, a) + {1\over v}d(a,b) + d(b,t)\}$, which means that
for going from a red to a blue point, the walkway is used from $a$ to $b$. So the set $\bigb (s,y)$ for a point $s\in R$ is either (see figure~\ref{unit}):

\begin{itemize}
\item the disk of radius $y$ centered at $s$ if $y<d(s,a)+{1\over v}d(a,b)$, or
\item the union of a disk of radius $y$ centered at $s$ and a disk of radius $y-d(s,a)-{1\over v}d(a,b)$ centered at $b$, otherwise.
\end{itemize}

To check the intersection condition, we proceed by first sorting the points in $R$ in nondecreasing order of their distance to $a$. We denote by $s_1, s_2, \ldots ,s_{|R|}$ the points in $R$, such that $d(s_1,a)\leq d(s_2,a)\leq \ldots \leq d(s_{|R|}, a)$. 

For each point $t\in B$, we identify, in $O(\log |R|)$ time, the smallest index $i$ such that $\forall j\geq i : d(s_j, a) > y - {1\over v}d(a,b) - d(b, t)$. All the red points $s_j$ for $j\geq i$ are such that we cannot go from $s_j$ to $t$ in time less than $y$ by using the walkway. So for all these points, we must be able to reach $t$ directly, which means that $t$ must lie in the intersection of the Euclidean unit disks ${\mathcal D} (s_j, y)$ with centers $s_j$ and radius $y$.

To check the latter condition, we can use a simple point location structure for disk intersections. The difficulty lies in the fact that we must keep a version of the point location structure for each intersection of the form $C_i=\bigcap_{j=i}^{|R|} {\mathcal D} (s_j, y)$ for $i=1,2,\ldots ,|R|$. To this purpose, we construct the intersections $C_i$ by adding the disks one by one in an incremental data structure. This incremental data structure can then be made persistent using known techniques~\cite{persistence}, with a $O(1)$ multiplicative cost.

The structure we use is made of two balanced search trees, one for the upper boundary of the intersection $C_i$, and another for the lower boundary. These trees allow to compute, in time logarithmic in the number of vertices of $C_i$, for any value of the $x$-coordinate, the highest (resp. lowest) point on the boundary of $C_i$ having this $x$-coordinate. From this, we can check if a point belongs to $C_i$. We now need a technical lemma.
\begin{lemma}
Consider the intersection $I$ of $i$ unit disks, and an additional, distinct, unit disk $\mathcal D$. The boundary of $\mathcal D$ intersects the boundary of $I$ in at most two points.
\end{lemma}
This shows that the number of vertices of $C_i$ is $O(i)$ and that the point location data structure answers queries in time $O(\log i) = O(\log n)$.
Using the partial-persistence transform~\cite{persistence}, the additional space cost due to persistence is the total number of pointer changes on the structure over a sequence of operations. Since there are at most two new vertices on the intersection of the disks for each new value of $i$, the total number of changes is $O(|R|)$. So the persistent data structure uses linear space.

So for each point $t\in B$, we can check in $O(\log n)$ time whether $t\in C_i$ for the computed index $i$. Hence the whole algorithm takes $O(n\log n)$ deterministic time.
\qed\end{proof}

\subsection{Minimum travel time diameter location of horizontal moving walkways}

Now given our algorithm for computing the travel time diameter and the observation that the 
location for $n$ source-destination pairs is LP-type, we are able to solve the minimum diameter 
location problem, which is in fact an implicit quasiconvex program.

\begin{prob}
\label{ap2Dhmww}
Given a set $P$ of $n$ points, and a real number $v>1$, 
find a horizontal moving walkway $[a^*,b^*]$ of speed $v$ minimizing the travel time diameter of $P$:
\begin{equation*}
\min_{a, b: a_y=b_y} \max_{s, t\in P} \dtime (s, t, a, b).
\end{equation*}
\end{prob}

\begin{theorem}
Problem~\ref{ap2Dhmww} can be solved in $O(n\log n)$ randomized expected time.
\end{theorem}
\begin{proof}
The result is by using lemma~\ref{chan2005}.
This requires to be able to decompose a set of points $P$ into $r$ groups $P_i$, each of size at most $\alpha |P|$, such that the set of constraints encoded by $P$ (the pairs $s,t\in P$) is the union of the corresponding sets for the $P_i$.
Here, we can partition the set $P$ in three equal-sized subsets, say $Q,R$ and $S$. We form the following three groups of points : $P_1 = Q\cup R$, $P_2=R\cup S$ and $P_3=Q\cup S$, each group containing $2n/3$ points. We have $r=3$ and $\alpha = 2/3$. This decomposition satisfies 
requirement~\ref{chansplit} of the lemma, because each pair of points is included in at least one group.

Another thing we need from requirement~\ref{chandec} of the lemma is an algorithm for the satisfaction-violation test for a group of constraints. This corresponds to a decision algorithm for the travel time diameter, which, from the previous lemma, can run in $O(n\log n)$ time. 

Hence from lemma~\ref{chan2005}, the optimization problem can be solved within the same time bounds, in the randomized expected sense. 
\qed\end{proof}

\subsection{An algorithm for moving walkway location in the plane}\label{subsec:approx}

So far we only considered the constrained problem in which the moving walkway must be horizontal. The unconstrained problem is as follows.

\begin{prob}
\label{ap2Dmww}
Given a set $P$ of $n$ points and a real number $v>1$, find a moving walkway $[a^*,b^*]$ of speed $v$ minimizing the travel time diameter of $P$:
\begin{equation*}
\min_{a, b} \max_{s,t\in P} \dtime (s, t, a, b).
\end{equation*}
\end{prob}

It is easy to verify that the constraints are not quasiconvex anymore, so we are not able to apply the previous randomized techniques.
We can solve the problem approximately, however, using the following trick: rotate the set of points in $O(1/\epsilon )$
directions, and for each direction solve the constrained version of the problem.

\begin{theorem}
A $(1+\epsilon )$-approximate solution for problem~\ref{ap2Dmww} can be found in $O(\frac{v}{\epsilon}n\log n)$ randomized expected time.
\end{theorem}
\begin{proof}
For each angle $i\epsilon /v$, with $i=0,1,\ldots , \lfloor 2\pi v / \epsilon \rfloor$, we rotate the set of points of this angle and
solve Problem~\ref{ap2Dhmww}. Among all the solutions computed, the one minimizing the diameter is retained.

The angle of the walkway returned by this procedure and the angle of an optimal moving walkway
$a^*, b^*$ differs by at most $\alpha = \epsilon /v$. Also, the returned solution is not worse than a walkway $a,b$ with the same orientation and such that 
$a=a^*$ and $d(a, b)=d(a^*, b^*)$. Then for any pair $s, t$ we have
\begin{eqnarray*}
d(s, a) + {1\over v}d(a, b) + d(b, t) & \leq & d(s, a^*) + {1\over v}d(a^*, b^*) + d(b^*, t) + d(b, b^*) \\
& \leq & d(s, a^*) + {1\over v}d(a^*, b^*) + d(b^*, t) + 2\sin \frac{\alpha}{2} d(a^*, b^*)\\
& \leq &  d(s, a^*) + {1\over v}d(a^*, b^*) + d(b^*, t) + \alpha d(a^*, b^*)\\
& \leq &  d(s, a^*) + {1\over v}d(a^*, b^*) + d(b^*, t) + {\epsilon\over v} d(a^*, b^*)\\
& \leq & (1+\epsilon ) ( d(s, a^*) + {1\over v}d(a^*, b^*) + d(b^*, t) ) .
\end{eqnarray*}
Note that the dependence on $v$ is necessary. In particular, if $v\to +\infty$, the above procedure is not applicable.
\qed\end{proof}

\begin{remark}
By combining the above trick with a grid technique (see for instance Har-Peled~\cite{HP01}), we can obtain a location algorithm whose complexity is of the 
form $O(n+{\text poly} (\epsilon^{-1}, v))$. The main idea is to snap the points in $P$ to a grid of size $\epsilon^{-1}\times \epsilon^{-1}$, and work 
on the point set formed by taking the centers of all nonempty cells. We can then run the above algorithm for $n=O(\epsilon^{-2})$.
\end{remark}

\begin{remark}
Finding an efficient exact algorithm for problem~\ref{ap2Dmww} is the natural extension of this work. It is easy to see that the problem is polynomial. In the unconstrained setting, every pair of points of the set defines a constraint consisting of pieces of $4$-variate functions in $5$-space. The pieces are algebraic functions of bounded degree. Thus, the complexity of the arrangement is polynomial, as well as the time to compute the upper envelope (see, e.g.,~\cite{daven}) and to find the minimum value over the envelope. 
Note however that the complexity of the arrangement is in $O(n^8)$, which means that this approach is probably not tight. Whether a better upper bound can be derived is left open.
\end{remark}

\section{Escalators and Elevators}
\label{sec:EaE}

In this section, we consider a number of variants of the moving walkway model, by combining a number of restrictions on the way the walkway is used.
One of the model we obtain is the {\em elevator} model, for which the location algorithm becomes linear. In the last part of this section, we give a simple algorithm for computing the travel time diameter of a set of points with respect to $k$ elevators.

\subsection{Variants}

We consider three restrictions to our model.

\paragraph{Unidirectional walkways.} In that case, we assume that the walkway can only be used for going from $a$ to $b$, and not for going from $b$ to $a$. If we simply assume this, the diameter of a point set is trivially the geometric diameter, since for each pair of points, the walkway cannot be useful in both directions, so the travel time in one of the two directions will be equal to the geometric distance between the points. To make the model interesting, we must further assume that the travel direction for each pair is prescribed. This can be achieved for instance by dividing the point set in two subsets, say red and blue points, and assuming that we always travel from a red point to a blue point. In this model, the constraints defining the problem are all quasiconvex, whatever the orientation of the walkway is. And since the diameter algorithm can be adapted, we immediately have a $O(n\log n)$ randomized expected time algorithm for the optimal location of a (unidirectional, red-blue) walkway of arbitrary orientation.

\paragraph{Mandatory use.} With this restriction, we assume that customers are obliged to use the walkway. In that case, the algorithm for computing the diameter boils down to computing the two points (one in each color set) that are the furthest away from $a$ and $b$, respectively. The whole location algorithm therefore becomes $O(n)$ for a horizontal walkway. If combined with the previous restriction, we obtain a $O(n)$ algorithm for the optimal location of a (mandatory, unidirectional, red-blue) walkway of arbitrary orientation. We call this model the {\em escalator} model, as the red and blue subsets can represent source and destination points in two different floors of a building.

\paragraph{Elevators.} The {\em elevator} model is obtained by assuming the previous two restrictions and letting $a=b$. In this case, the problem is similar to a point location problem mentioned by Eppstein (\cite{qconvex}, section 3.2), and solvable in $O(n)$ expected time using Chan's method again.

\subsection{Computing the travel time diameter with respect to $k$ elevators}

We now give an algorithm that, given the location of $k$ elevators, efficiently computes the diametral source-destination pair.
This can be considered as a first step towards an algorithm that would locate $k$ elevators instead of one.

\begin{prob}
\label{diamel}
Given a set $R$ of red points, a set $B$ of blue points, and a set $\{ e_i\}_{i=1,2,\ldots ,k}$ of points (elevators), find a pair $r,b$ solving
the following problem:
\begin{equation*}
\max_{r\in R, b\in B} \min_i d(r,e_i)+d(e_i,b).
\end{equation*}
\end{prob}

We assume that $|R|=|B|=n$.

\begin{theorem}
Problem~\ref{diamel} can be solved in $O(kn\log ^{k-1} n)$ deterministic time.
\end{theorem}
\begin{proof}
Suppose that $R=\{ r_i\}_{i=1,2,\ldots ,n}$ and $B=\{ b_i\}_{i=1,2,\ldots ,n}$. Elevator $e_1$ is used to go from $r_i$ to $b_j$ iff:
\begin{eqnarray}
d(r_i,e_1) + d(e_1,b_j) & \leq &  d(r_i,e_{\ell}) + d(e_{\ell},b_j)\ \forall \ell = 1,2,\ldots ,k \\
\underbrace{d(r_i,e_1) - d(r_i,e_{\ell})}_{:=P^{(i)}_{\ell}} & \leq & \underbrace{d(e_{\ell},b_j) -  d(e_1,b_j)}_{:=Q^{(j)}_{\ell}}\ \forall \ell = 1,2,\ldots ,k .
\end{eqnarray}
After having computed the vectors $P^{(i)}$ and $Q^{(j)}$ for each $r_i$ and each $b_j$ respectively, finding
the maximum time distance between two points using elevator $e_1$ amounts to finding the maximum distance 
$d(r_i,e_1)+d(e_1,b_j)$ between all pairs $r_i,b_j$ for which $Q^{(j)}$ dominates $P^{(i)}$. This can be done in
$O(n\log ^{k-1} n)$ using a divide-and-conquer algorithm. Doing this once for each elevator yields the result.\qed
\end{proof}

\bibliography{walkwaysandelevators.bib}
\bibliographystyle{abbrv}

\end{document}